# POWDER DIFFRACTION DATA AND MESOMORPHIC PROPERTIES FOR 4-BUTYLOXYPHENYL 4'-DECYLOXYBENZOATE

**M. Cvetinov[1], M. Stojanović[1], D. Obadović[1], S. Rakić[1], A. Vajda[2], K. Fodor-Csorba[2], N. Éber[2]**

[1] Department of Physics, Faculty of Sciences, University of Novi Sad, Novi Sad, Serbia
[2] Institute for Solid State Physics and Optics, Wigner Research Centre for Physics, Hungarian Academy of Sciences, Budapest, Hungary

corresponding author:
Miroslav Cvetinov
Trg Dositeja Obradovića 4
21000 Novi Sad
Phone number: + 381(21)485-28-24
email: miroslav.cvetinov@df.uns.ac.rs

Maja Stojanović, email: maja.stojanovic@df.uns.ac.rs
Dušanka Obadović : dusanka.obadovic@df.uns.ac.rs
Srđan Rakić: srdjan.rakic@df.uns.ac.rs
Aniko Vajda: vajda@szfki.hu
Katalin Fodor-Csorba: fodor@szfki.hu
Nandor Eber: eber@szfki.hu

**Abstract**

Unit cell parameters obtained from X-ray powder diffraction data are presented for the crystalline phase of a liquid crystal 4-butyloxyphenyl 4'-decyloxybenzoate: $a = 23.098\,(4)$ Å, $b = 5.974\,(6)$ Å, $c = 12.357\,(10)$ Å, $\beta = 121.53\,(8)°$, unit-cell volume $V = 1453.56$ Å$^3$. Temperature dependent X-ray diffraction data confirmed the existence of smectic A and smectic C mesophases and a more ordered, tilted crystalline smectic phase. Possibility of existence of previously reported smectic B phase as well as another crystalline phase was refuted.

# I. INTRODUCTION

In order to respond to the application requirements of the liquid crystalline display industry, tens of thousands of different liquid crystalline substances have been synthesized up to now. Widespread technologies for display operation are highly reliant on different properties of nematic liquid crystalline phase, like its temperature range, its closeness to the room temperature, viscosity, conductivity, dielectric and optical anisotropies [1]. Moreover, smectic phases have found commercial application in another niche market of display industry [2]. Modern multicomponent mixtures that are most often used for industrial applications, are highly dependaent on polymorphism and molecular shapes and properties of starting compound [3, 4]. Thus, full characterisation of liquid crystalline substances including crystallographic data and molecular modelling calculations are of interest for the industry [5-7].

4,4'-alkyl/alkoxyphenylbenzoates represent known for a long time huge family of calamitic liquid crystals. They are prototypal liquid crystals; more compounds of their homologous series have been synthesized than of any other series with a single core unit [8]. Miscibility studies were performed on them in order to adjust their physical properties and to induce or suppress specific mesophases [9]. Precursors of these compounds, oxybenzoic acids, and their polymers were also studied by the means of X-rays powder diffraction as mesophases and solid phases [10, 11]. In addition, polymerized forms of oxybenzoates, poly(p-oxybenzoate) were characterized by combined means of electron diffraction and molecular modelling [12].

Despite of this fact, crystallographic data for these types of compounds remain scarce. Inspecting the literature structural information was found only for 4-octyloxyphenyl 4'-pentyloxybenzoate (**5OO8**), which was elucidated using X-ray diffraction on single crystal sample [13]. The crystalline phase of this compound is triclinic with unit cell parameters: $a = 5.6132(5)$ Å, $b = 13.5726(11)$ Å, $c = 15.9505(13)$ Å, $\alpha = 77.0400(10)°$, $\beta = 83.3950(10)°$, $\gamma = 88.4110(10)°$ and unit cell volume $V = 1176.38(17)$ Å$^3$.

In the present paper we report on our X-ray and molecular modelling studies on the title compound, 4-butyloxyphenyl 4'-decyloxybenzoate (**10OO4**) which is another member of the 4-4'-alkoxyphenylbenzoates homologous series. It differs from **5OO8** only in the lengths of the alkyl chains at both ends of the molecule. The investigations have been carried out in the crystalline as well as in the liquid crystalline phases.

# II. MATERIALS AND METHODS

For the synthesis of the title compound the starting materials were purchased from Aldrich and used without further purification. During the synthesis 4-decyloxy benzoic acid was converted into its acid chloride by the thionylchloride technique and it was reacted with 4-butiloxyphenol in the presence of triethyl amine. The chemical formula of the resulting **10OO4** is shown in the Figure 1.

According to the literature [14], the studied compound exhibits isotropic liquid (I), enantiotropic nematic (N), smectic A (SmA), smectic C (SmC), smectic B (SmB) and crystalline (Cr) phases, as shown in Figure 2. We note, however, that there is no full agreement on the polymorphism of **10OO4** in the literature [8, 14, 15].

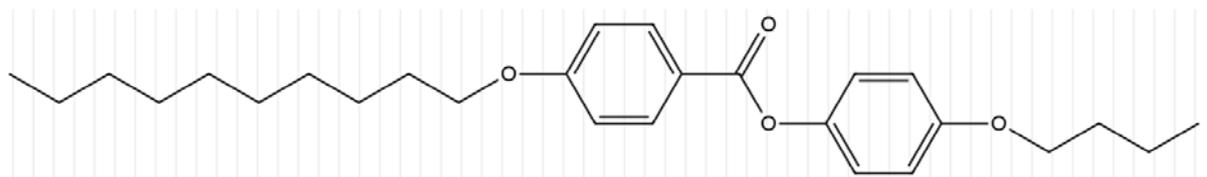

Figure 1. Chemical formula of **10OO4.**

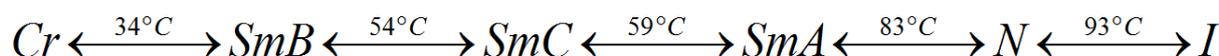

Figure 2. Phase sequence of **10OO4** [14].

The diffraction pattern for the title compound was recorded in the crystalline phase at room temperature with a Seifert V-14 powder diffractometer equipped with a goniometer in the Bragg–Brentano ($\theta$:$2\theta$) geometry, using CuK$\alpha$ radiation without monochromator ($\lambda$CuK$\alpha_1$ = 1.5406 Å, Ni filter, generator setting: 30 kV, 30 mA). Indium(III) oxide (In$_2$O$_3$) was used as an external standard. XRD data were collected over the $2\theta$ range from 4° to 50° with a step size of 0.02° and a counting time of 4 s per step.

Non-oriented samples were also investigated in the mesophases of **10OO4** at different temperatures on another channel of the same diffractometer, which is equipped with an automated high-temperature kit Paar HTK-10. Continuous $2\theta$ scan with the speed of 2° per minute was employed. Due to the fact that raw data were used (unfitted and without CuK$\alpha_2$ stripping), the temperature dependent $d$-values were calculated using the averaged wavelength of CuK$\alpha$ radiation ($\lambda$ = 1.5418 Å). Molecular parameters were calculated by RM1 semi-empirical quantum chemistry level [21].

## III. RESULTS AND DISCUSSION

The experimental powder diffraction pattern is depicted in Figure 3. Profile fitting procedure was used to determine the peak positions and intensities of the diffraction peaks; the software X-Fit [16] was employed for this purpose.

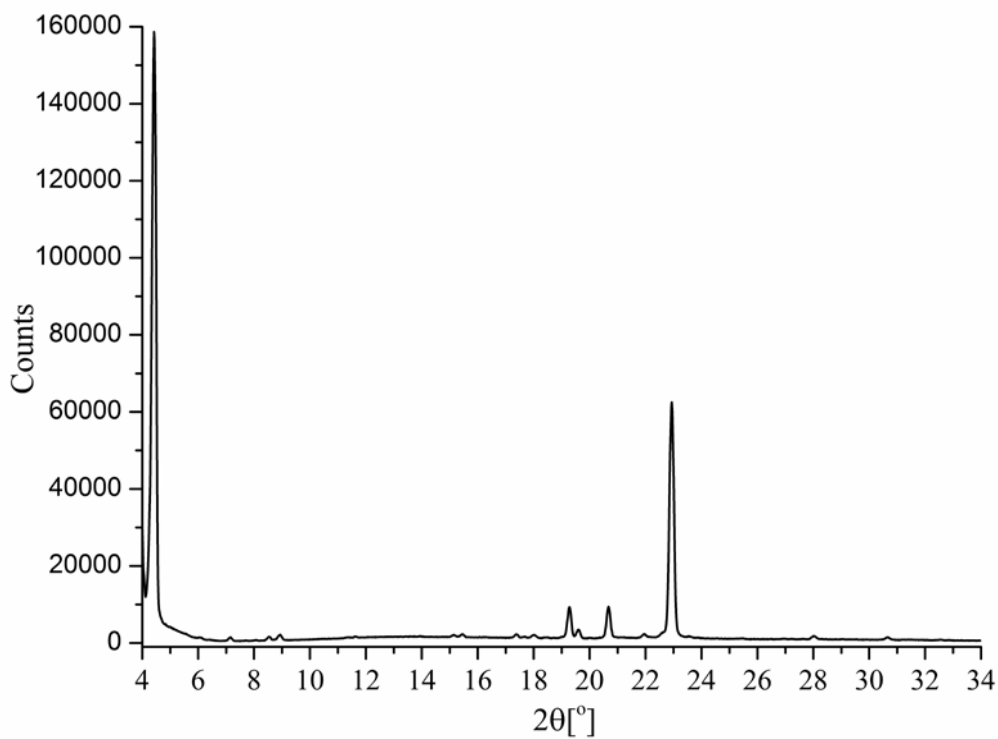

Figure 3. X-ray powder diffraction pattern of **10OO4** at room temperature.

The peak locations ($2\theta_{obs}$), the relative peak intensities ($I_{rel}$) and the corresponding $d$-values ($d_{obs}$) calculated using Bragg's law based on the wavelength of the Cu$K\alpha_1$ radiation ($\lambda = 1.5406$ Å) are listed in Table I. After correction for sample displacement and zero shift errors, the unit cell parameters were derived using TREOR90 [17] within the CRYSFIRE frontend application [18] with the results all being within the errors indicated. Unit-cell parameters were found to be: $a = 23.098$ (4) Å, $b = 5.974$ (6) Å, $c = 12.357$ (10) Å, $\beta = 121.53$ (8)°, unit-cell volume $V = 1453.56$ Å$^3$. The following figures of merit were achieved $F_{15} = 11$ (0.017550, 78) [19] and $M_{15} = 12$ [20]. Table I also lists the diffraction peak locations ($2\theta_{calc}$) and the corresponding $d$-values ($d_{calc}$) calculated from the cell parameters given above which agree well with the observed values.

Table I. Indexed X-ray powder diffraction data for 10OO4. Only peaks with $I_{rel} \geq 0.3$ are presented.

| $2\theta_{obs}$ | $d_{obs}$ | $I_{rel}$ | h | k | l | $2\theta_{calc}$ | $d_{calc}$ | $\Delta 2\theta$ |
|---:|---:|---:|---:|---:|---:|---:|---:|---:|
| 4.464 | 19.7781 | 100.00000 | 1 | 0 | 0 | 4.484 | 19.69054599 | -0.020 |
| 7.162 | 12.3326 | 0.49209 | -1 | 0 | 1 | 7.149 | 12.35517742 | 0.013 |
| 8.547 | 10.3371 | 0.51837 | -2 | 0 | 1 | 8.495 | 10.4003273 | 0.052 |
| 8.935 | 9.8890 | 0.89257 | 2 | 0 | 0 | 8.976 | 9.844051711 | -0.041 |
| 15.191 | 5.8278 | 0.61885 | 2 | 0 | 1 | 15.184 | 5.830394526 | 0.007 |
| 15.466 | 5.7247 | 0.35219 | 1 | 1 | 0 | 15.487 | 5.716999157 | -0.021 |
| 17.356 | 5.1053 | 0.34277 | 2 | 1 | 0 | 17.349 | 5.107383448 | 0.007 |
| 18.018 | 4.9192 | 0.30332 | 4 | 0 | 0 | 18.007 | 4.922208309 | 0.011 |
| 19.275 | 4.6012 | 4.62320 | 3 | 0 | 1 | 19.300 | 4.595263453 | -0.025 |
| 19.598 | 4.5261 | 0.99432 | 1 | 0 | 2 | 19.578 | 4.530635164 | 0.020 |
| 20.675 | 4.2927 | 4.79861 | -2 | 1 | 2 | 20.665 | 4.294707903 | 0.010 |
| 21.953 | 4.0456 | 0.79782 | -2 | 0 | 3 | 21.968 | 4.042829627 | -0.015 |
| 22.927 | 3.8759 | 39.80727 | 2 | 0 | 2 | 22.923 | 3.876514624 | 0.004 |
| 28.016 | 3.1823 | 0.84121 | 5 | 0 | 1 | 28.012 | 3.182748741 | 0.004 |
| 30.657 | 2.9139 | 0.72388 | 4 | 0 | 2 | 30.643 | 2.915208305 | 0.014 |

Inspecting the unit cell parameters one finds that the crystal system of **10OO4** is different from that of the closely related compound, **5OO8**. Moreover, the unit-cell volume of **10OO4** is about 19% larger than that of **5OO8**. This difference probably stems from the asymmetry of molecules around their ester groups and the different lengths of molecules themselves. According to molecular modeling, **10OO4** has a length of 28.72 Å between the carbon atoms at its both ends, while **5OO8** has a length of 27.37 Å. Length of rigid cores of both molecules is 11.72 Å measured between oxygen atoms attached to phenyl benzoate moiety. The angle between the two neighboring phenyl rings in the molecular core is 44.15°. The electrostatic potential map of compound **10OO4** is shown in Figure 4. The density of electrons appears as colours changing from blue to red; it is highest in the vicinity of atoms characterized by high electronegativity (principally at oxygen atoms).

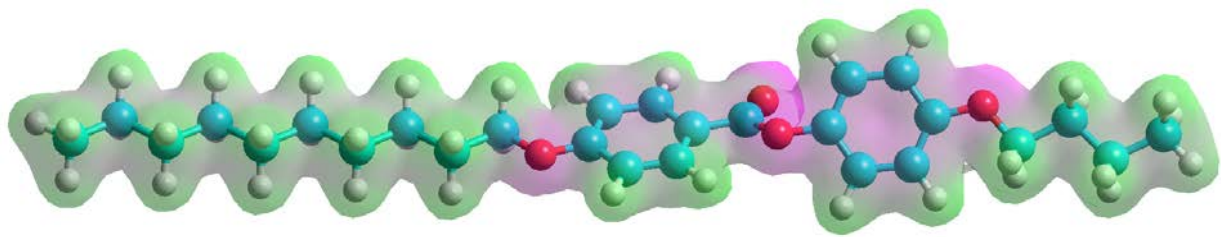

Figure 4. Electrostatic potential map of the minimum energy conformation of **10OO4**

In Figure 5 we present typical diffraction spectra of **10OO4** for a sequence of temperatures, in order to demonstrate the changes occurring at the phase transitions. SmA phase is indicated by the small angle reflection at $2\theta = 3.50°$, corresponding to the thickness of the smectic layers $d = 25.22$ Å; the broad large angle scattering is due to the fluidic order within the smectic layers and represents the average intermolecular distance between the long axes of neighboring molecules [22, 23]. For the simplest SmA phase, the layer spacing calculated from the Bragg peak is slightly less than the molecular length. This is found for systems with a little tendency for molecular association, and hence a complete head-to-tail disorder [24]. Another small angle reflection emerges at $2\theta = 3.86°$ at 75 °C in cooling, indicating a reduction of the thickness of smectic layers to $d = 22.89$ Å, which corresponds to a tilt angle of about 37° of the molecules in the SmC phase.

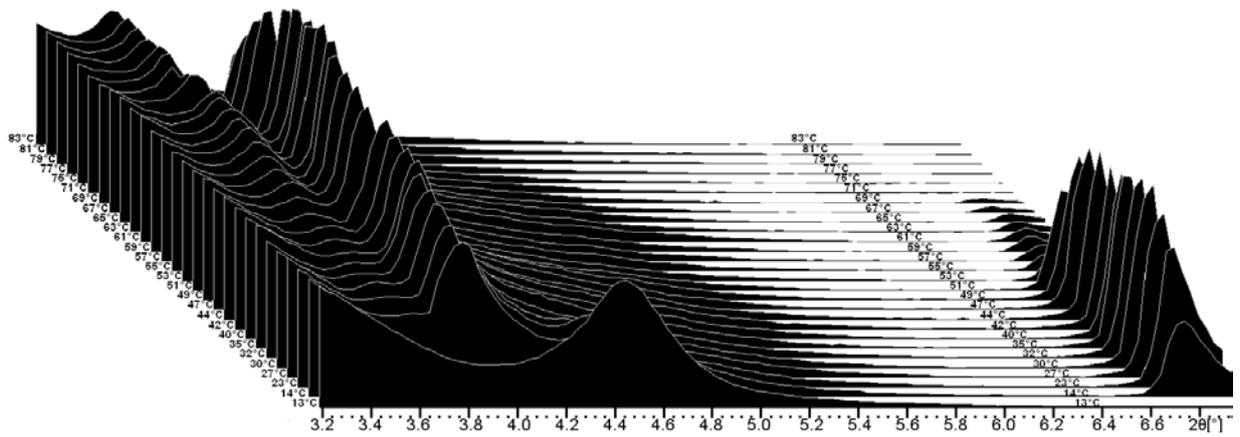

Figure 5. X-ray diffractograms of **10OO4** in the angle range $3.2° < 2\theta < 7°$ at different temperatures obtained in cooling.

The X-ray diffraction study confirmed the existence of a more ordered smectic phase, which is primarily characterized by a first order reflection peak that appears at $2\theta = 6.76°$ (Figure 5 and 6) in cooling below 53 °C. Both the existence of sharp peaks indicative of long range in-plane order and the presence of increased scattering in the proximity of $2\theta = 20°$ (Figure 6) indicative of hexatic short range in-plane structures, strongly suggest that this more ordered phase below 53 °C is one of the crystalline smectic phases (CrX). The use of aligned (monodomain) sample would be necessary to clarify the long-range and short-range positional order more precisely and thus to resolve clearly the nature of the CrX phase. Note that the phase at this temperature range was assigned as SmB in [14]. This is, however, highly improbable because the small angle peak at $2\theta = 3.86°$ remained in place during the SmC–CrX phase transition and even down to 14 °C, although with a lower intensity. The unchanged layer spacing therefore limits the possible types of the CrX phase to one of the tilted crystalline smectic phases (CrG, CrJ, CrH or CrK) [24].

Finally, the CrX-Cr phase transition was difficult to test in cooling; the reason is the supercooling tendency of the CrX phase. To obtain a stable crystalline phase with the diffraction pattern as in Figure 3 and Table I, the sample was left for 1 day at 13 °C.

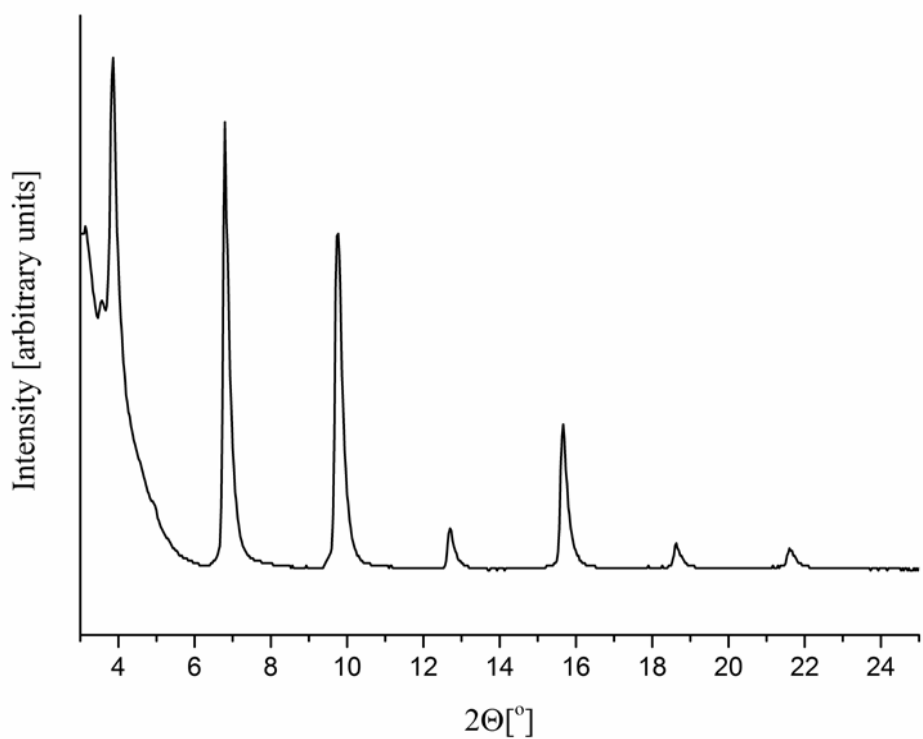

Figure 6. X-ray powder diffraction pattern of the more ordered smectic phase of **10OO4** at 40°C using Cu*Kα* radiation.

## IV. CONCLUSION

Temperature dependence of microstructural characteristics of the liquid crystal **10OO4** was studied by powder X-ray diffraction. The rich polymorphism of the compound was confirmed. On cooling through the SmC temperature range, layer spacing did not show any consistent trend toward either increasing or decreasing ($d = 22.89$ Å $\pm 0.06$ Å in the whole SmC range). At low temperatures an ordered mesophase of yet unknown nature was found, which could be one of the tilted crystalline smectic phases. At room temperature the compound takes a monoclinic crystalline phase, whose lattice parameters could be determined.

**ACKNOWLEDGMENTS**

This work was partly supported by the research Grant No. OI171015 from the Ministry of Education and Science of the Republic of Serbia, the Hungarian Research Fund OTKA K81250 and the SASA-HAS bilateral scientific exchange project #9.